\documentclass[prd, preprint, aps, showpacs]{revtex4}

\usepackage{graphicx}
\usepackage{color}

\def\be{\begin{equation}}
\def\ee{\end{equation}}
\def\ba{\begin{eqnarray}}
\def\ea{\end{eqnarray}}
\def\beq{\begin{eqnarray}}
\def\eeq{\end{eqnarray}  }

\def\eref#1{Eq.~(\ref{#1})}

\def\rmd{{\rm d}}

\def\lp{\left(}
\def\rp{\right)}

\begin{document}
\title{Extended Lifetime in Computational
Evolution of Isolated Black Holes}

\author{Matthew Anderson}
\author{Richard A. Matzner}
\affiliation{Center for Relativity, University of Texas at Austin,
Austin, TX 78712-1081, USA}

\begin{abstract} Solving the 4-d Einstein equations as evolution
in time requires solving equations of two types: the four elliptic initial
data (constraint) equations, followed by the six second order evolution
equations. Analytically the constraint equations remain solved under the
action of the evolution, and one approach is to simply monitor them ({\it
unconstrained} evolution).

The problem of the 3-d computational simulation of even a single
isolated vacuum black hole has proven to be remarkably difficult.
Recently, we have become aware of two publications that describe
very long term evolution, at least for single isolated black
holes.
An essential feature in each of these results is {\it constraint
subtraction}. Additionally, each of these approaches is based on what we
call ``modern," hyperbolic
formulations of the Einstein equations. It is generally assumed, based on
computational experience, that the use of such modern formulations is
essential for long-term black hole stability. We report here on comparable
lifetime results based on the much simpler (``traditional") $\dot g$ - $\dot
K$ formulation.

With specific subtraction of constraints, with a simple
analytic gauge, with very simple
boundary conditions, and for moderately
large domains with moderately fine resolution, we find computational
evolutions of isolated nonspinning black holes for times exceeding $1000
GM/c^2$.

We have also carried out a
series of {\it constrained } 3-d evolutions of single isolated
black holes. We find that constraint solution can produce
substantially stabilized long-term single hole evolutions.
However, we have found that for large domains, neither constraint-subtracted
nor constrained  $\dot g$ - $\dot K$
evolutions carried out in Cartesian coordinates admit arbitrarily
long-lived simulations. The failure appears to arise from features at the
inner excision boundary; the behavior does generally improve with
resolution.

\end{abstract}

\pacs{ }

\maketitle

%----------------------------------------------------------------------
%
%
%
%----------------------------------------------------------------------
\section{Introduction}
\label{sec:intro}

Binary black hole systems are expected to be the strongest possible
astrophysical gravitational wave sources. In the final moments of
stellar mass black hole inspiral, the radiation will be detectable in
the current (LIGO-class) detectors. If the total binary mass is of the
order of $10M_{\odot}$, the moment of final plunge to coalescence will
emit a signal detectable by the current generation of detectors from
very distant (Gpc) sources. The merger of supermassive black holes in
the center of galaxies will be the dominant signal in the
spaceborne LISA detector, and detectable out to large redshift.

Simulation of these mergers will play an important part in the
prediction, detection, and the analysis of their gravitational signals
in gravitational wave detectors. To do so requires a correct formalism which
does not generate spurious singularities during the attempted simulation.
Recent important work\cite{schKidteuk}\cite{yo} has been
done in extending the computational lifetime
of single isolated black hole simulations. We report here on such an
extension
which demonstrates that {\it constraint subtraction} by itself is adequate
to produce very long-lived simulations. We demonstrate this even for a very
simple (``traditional") $\dot g$ - $\dot
K$ formulation, with specific subtraction of constraints (which
are analytically zero) for single isolated nonspinning
black holes, with a simple
analytic gauge (lapse and shift {\it not} ``densitized";
see \ref{sec:densitized} below), with very simple
boundary conditions, and for moderately
large domains with moderately fine resolution.

We have also found that {\it constrained }3-d evolution {\it with}
densitized lapse can produce
substantially stabilized long-term single holes, even for subtractions
that
differ from the values that we have found to be optimal in the
unconstrained
case, and we give some preliminary constrained evolution results.

However, in all cases we find that attempting to carry out the simulations
on very large domains ($\pm 20M$, or larger) still yields eventually
unstable simulations, by any of the methods reported here.

\section{$3+1$ Formulation of Einstein Equations}

We take a Cauchy formulation
(3+1) of the ADM type, after Arnowitt, Deser, and
Misner~\cite{ADM}. In such a method the 3-metric $g_{ij}$ and its momentum
$K_{ij}$ are specified at one initial time on
a spacelike hypersurface, and evolved into the future.  The ADM metric is
\be
\rmd s^2 = -(\alpha^2 - \beta_i \beta^i)\,\rmd t^2 + 2\beta_i \, \rmd t
\,\rmd x^i
     + g_{ij}\, \rmd x^i\, \rmd x^j
\label{eq:admMetric}
\ee
where $\alpha$ is the lapse function and $\beta^i$ is the shift
3-vector; these gauge
functions encode the
coordinatization.\renewcommand{\thefootnote}{\fnsymbol{footnote}}\setcounter
{footnote}{2}\fnsymbol{footnote}

\footnotetext[2]{Latin indices run $1,2,3$ and are lowered and raised
by $ g_{ij}$
and its 3-d inverse $ g^{ij}$.}
%\footnote{Latin indices run $1,2,3$ and are lowered and raised by $ g_{ij}$
%and its 3-d inverse $ g^{ij}$.}

The Einstein field equations contain both hyperbolic evolution equations
and elliptic constraint equations.
The constraint equations for vacuum in the ADM decomposition are:
\beq
H = \frac{1}{2} [R - K_{ij}K^{ij} + K^2] &=& 0,
\label{eq:constraintH}
\eeq
\beq
H^i = \nabla_j \lp K^{ij} - g^{ij}K\rp  &=& 0.
\label{eq:constraintK}
\eeq
Here $R$ is the 3-d Ricci scalar constructed from the 3-metric, and
$\nabla_j$ is the
3-d covariant derivative compatible with $ g_{ij}$.
Initial data must satisfy these constraint
equations; one may not freely specify all components of $g_{ij}$ and
$K_{ij}$.   

In this paper we are concerned only with single isolated black holes.
From this point of view the problem is not
solving the initial value equations, since the data are known analytically.
Instead, the question is one of the stability of the solution
as these data are evolved computationally.
The evolution equations from the
Einstein system are

\begin{equation}
       \dot g_{ij} = -2\alpha K_{ij} +\nabla_j\beta_{i} + \nabla_i\beta_{j}
\label{eq:gdot}
\end{equation}
and 

\begin{equation}
      \dot  K_{ij} =  - \nabla_i \nabla_j \alpha + \alpha ( R_{i j}
          -2K_{i k}K^k_{~j} + KK_{i j} ) + \beta^k \nabla_k K_{ij} + K_{ik} \nabla_j \beta^k + K_{jk} \nabla_i \beta^k
\label{eq:kdot}
\end{equation}
where a dot ( $\dot{}$ ) denotes the partial derivative with respect
to time, and $R_{ij}$ is the 3-d Ricci tensor.

We call this form of the Einstein equations {\it of ADM type},
referring to the fundamental development \cite{ADM}; this specific form is
called the $\dot g$ - $\dot K$ form. Here,
\eref{eq:constraintH}--\eref{eq:constraintK}, the
constraint equations, are the vacuum Einstein equations ${}^4 G_{00} = 0$
and
${}^4 G_{0i} = 0$ respectively. \eref{eq:gdot}--\eref{eq:kdot}, the
evolution equations, are a first order form of the vacuum Einstein equations
${}^4 R_{ij} = 0$.

The true ADM form writes the evolution equations as
${}^4 G_{ij} = 0$,
the space components of the 4-d Einstein tensor, rather than the
Ricci tensor. Frittelli and Reula\cite{49Frittelli} have shown that with
certain
(rather strong) assumptions, there is stable maintenance of the constraints
under unconstrained evolution for \eref{eq:gdot}--\eref{eq:kdot},
but only neutral stability for ${}^4 G_{ij} = 0$.

\section{Data Form}
\label{sec:DataForm}

In this paper we consider only single isolated black holes,
so the data setting problem is already solved; we use Kerr-Schild data,
which describes a single isolated spinning or nonspinning black hole
hole.

The Kerr-Schild~\cite{KerrSchild} form of a black hole solution describes
the 
spacetime of a single black hole
with mass, $m$, and specific angular momentum, $a = j/m$, in a coordinate
system that is well behaved at the horizon:
\be
        \rmd s^{2} = \eta_{\mu \nu}\,\rmd x^{\mu}\, \rmd x^{\nu}
                 + 2H_{KS}(x^{\alpha}) l_{\mu} l_{\nu}\,\rmd x^{\mu}\,\rmd
x^{\nu},
        \label{eq:1}
\ee
where $\eta_{\mu \nu}$ is the metric of flat space, $H_{KS}$ is a scalar
function of $x^\mu$, and $l_{\mu}$ is an (ingoing) null vector, null
with respect to both the background and the full metric,
\be
\eta^{\mu \nu} l_{\mu} l_{\nu} = g^{\mu \nu} l_{\mu} l_{\nu} = 0.
\label{eq:2}
\ee

Comparing the Kerr-Schild metric with the ADM
decomposition~\eref{eq:admMetric}, we find that the $t=\hbox{\rm constant}$
3-space metric is: $g_{ij} = \delta_{ij} + 2 H_{KS} l_i l_j$.
Further, the ADM gauge variables  are
\be
\beta_i = 2 H_{KS} l_0 l_i,
\ee
 and
\be
\alpha = \frac{1}{\sqrt{1 + 2 H_{KS} l_0^2}}.
\label{eq:ksAlpha}
\ee

The extrinsic curvature can be computed from Eq.(\ref{eq:gdot}):

\be
        K_{ij} = \frac{1}{2\alpha}[\nabla_j\beta_{i} + \nabla_i\beta_{j}
                     - \dot g_{ij}],
\label{eq:k_ks}
\ee

Each term on the right hand side of this equation is known analytically; in
particular, for a black hole at rest, $\dot g_{ij}=0$.

The general non-moving
black hole metric in Kerr-Schild form (written in Kerr's original
rectangular coordinates) has
\begin{equation}
        H_{KS} = \frac{mr}{r^{2} + a^{2}\cos^{2} \theta},
        \label{eq:ks_h}
\end{equation}
and
\begin{equation}
        l_{\mu} = \left(1, \frac{rx + ay}{r^{2} + a^{2}}, \frac{ry -
        ax}{r^{2} + a^{2}}, \frac{z}{r}\right),
        \label{eq:4}
\end{equation}
where $r,~ \theta$ (and $\phi$)
are auxiliary spheroidal coordinates,  $z = r(x,y,z) \cos \theta$,
and $\phi$ is 
the axial angle.  $r(x, y, z)$ is obtained from the relation,
\begin{equation}
        \frac{x^{2} + y^{2}}{r^{2} + a^{2}} + \frac{z^{2}}{r^{2}} = 1,
        \label{eq:5}
\end{equation}
giving
\begin{equation}
        r^{2} = \frac{1}{2}(\rho^{2} - a^{2}) +
        \sqrt{\frac{1}{4}(\rho^{2} - a^{2})^{2} + a^{2}z^{2}},
        \label{eq:6}
\end{equation}
with
\begin{equation}
\rho = \sqrt{x^{2} + y^{2} + z^{2}}.
        \label{eq:rho_def}
\end{equation}

In the
nonspinning case, one has $l_i= x_i/r$, so that
$\alpha = \frac{1}{\sqrt{1 + 2 m/r}}$, and $\beta_i = 2 m x_i/r$.

%\begin{figure}
%\begin{center}
%\includegraphics[width=3.5in,angle=270]{ham.pdf}
%\end{center}
%\caption{The Hamiltonian constraint (units $m^{-2}$) calculated for .}
%\label{fig:hc_with_att}
%\end{figure}

\section{Constraint Subtraction}
\label{sec:C-S}

\begin{figure}
\begin{center}
\includegraphics[width=5.0in, angle=0]{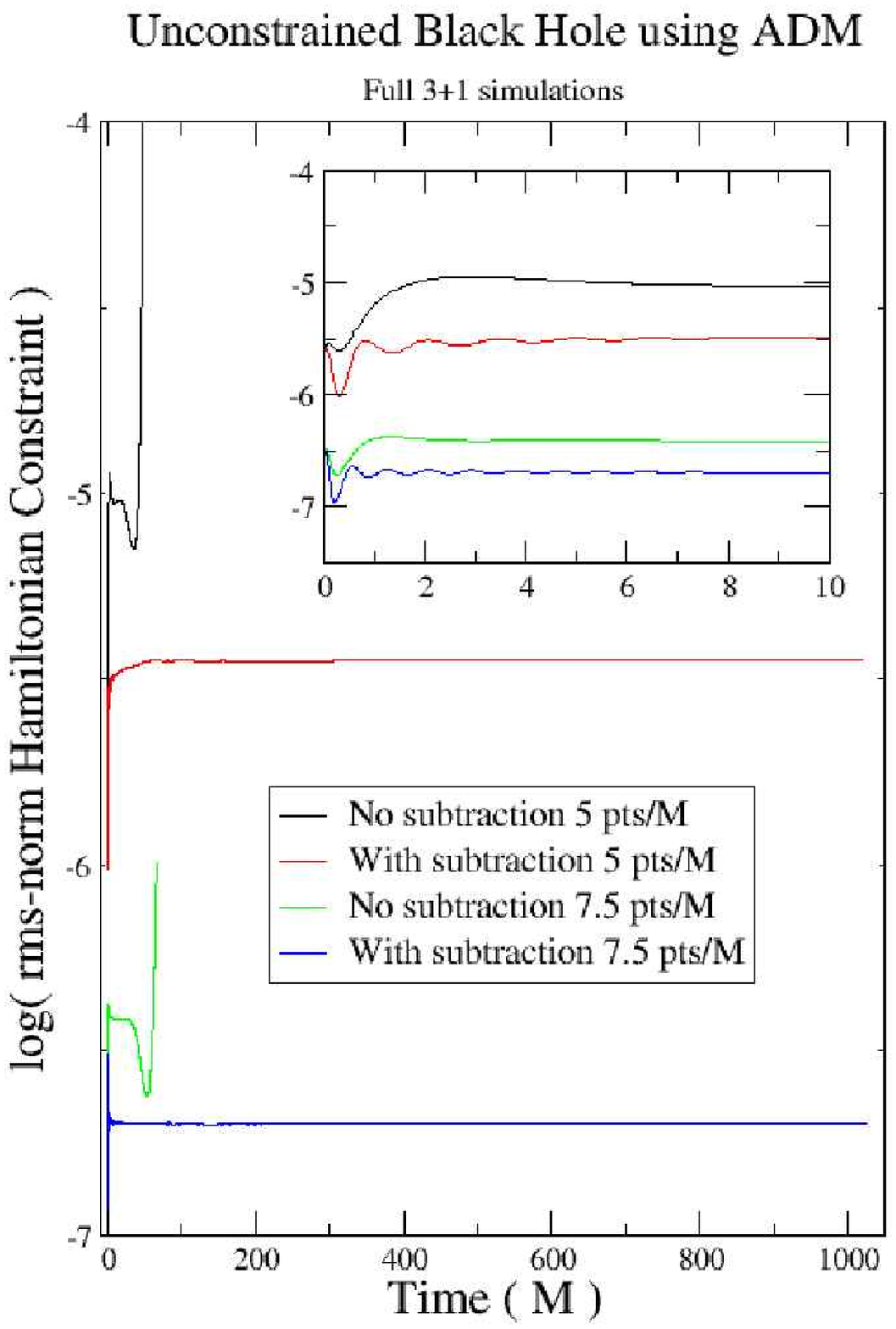}
\end{center}
\caption{The log of the rms norm of the Hamiltonian
constraint violations for
constraint-subtracted and unsubtracted nonspinning black hole simulations
with excision.  The simulations were performed at resolutions
of $M/5$ and $M/7.5$ on a domain size of $\pm 10M$.  The long-lived runs
employed optimal constraint subtraction (\eref{eq:constraintSub}). The
short-lived run
employed no subtraction.
}
\label{fig:standard_domain}
\end{figure}

\begin{figure}
\begin{center}
\includegraphics[width=5.0in, angle=0]{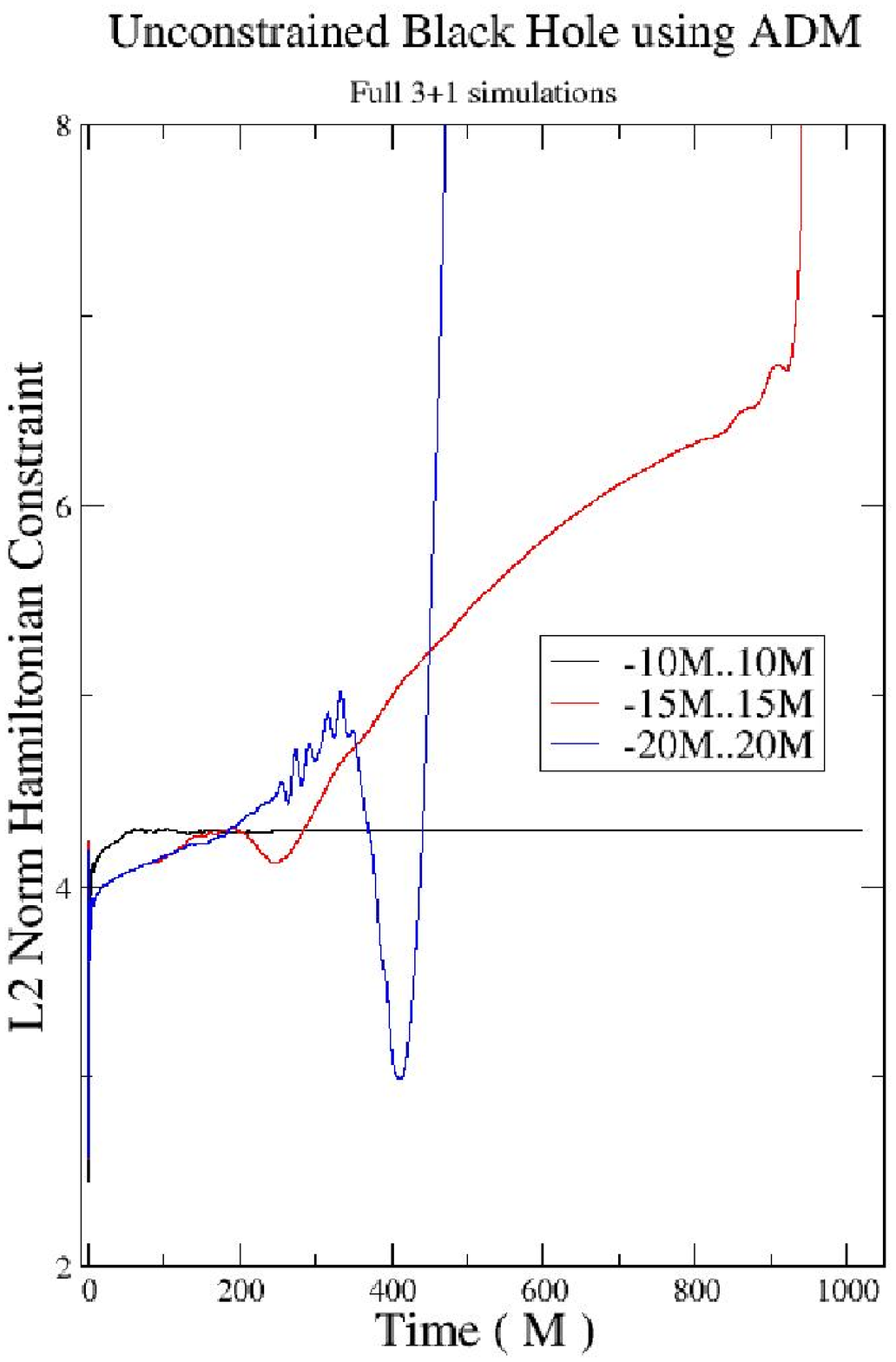}
\end{center}
\caption{The $l_2$ norm of the Hamiltonian constraint
violation for 
constraint-subtracted and unsubtracted nonspinning black hole
simulations with excision
performed at a resolution
of $M/5$ on a domain sizes of $\pm 10M$, $\pm 15M$, or $\pm 20M$. All cases
employed the optimal constraint subtraction (\eref{eq:constraintSub}).
As the computational domain size increases, the simulation is increasingly shorter lived.
}
\label{fig:vary_domain}
\end{figure}

The difference between \eref{eq:gdot}--\eref{eq:kdot},
and ${}^4 G_{ij} = 0$, is a specific subtraction of
the constraint equations. This has led to the
consideration by a number of groups, of constraint subtraction with
coefficients chosen by numerical search, or by analytical estimate (perhaps
combined with numerical search) to improve the long-term stability of the
unconstrained evolution. We have carried out such a numerical search, and
we use the following constraint subtraction:

\be
-\alpha H (0.464\, g_{i\,j}\, + 0.36\ K_{i\,j}\,)
\label{eq:constraintSub}
\ee
on the right hand side of the $\dot K_{ij}$ equation (\eref{eq:kdot}).
We have found that this subtraction
substantially improves the unconstrained evolution
of nonspinning single-hole data. For these evolutions we choose fixed
(Dirichlet) outer
boundary conditions set equal to the analytical value.  In every case the
simulation excises the interior of the black hole.
One-sided differencing is used near the inner (mask) boundary, so no
boundary condition is needed there, consistent with the property of the
horizon as a causal boundary. The mask is specified at a radius of 0.75 M.
The
solution employs the SUNDIALS package for time integration \cite{SUNDIALS,CVODES}.
The spatial
discretization is fourth order, and the
time evolution is typically fourth order (variable-order according
to the relative and absolute error
tolerances).  This approach successfully stabilizes the evolutions for
domains of $\pm 10M$, as shown in Figure \ref{fig:standard_domain}.
For larger domains, however, the system is increasingly shorter lived, as shown
in Figure \ref{fig:vary_domain}.
As a simple measure of the quality of solution, we present the either the
$l_2$ norm or rms norm of the Hamiltonian constraint constructed via a
straightforward fourth order differencing scheme from the code results.

Although it is difficult to compare subtraction techniques across
different formal representations of the Einstein equations, we do find
typically much smaller coefficients of subtraction (of order $0.5$) than
found for different formulations, e.g. \cite{schKidteuk} with a constraint
subtraction of order $-12.$ The BSSN formulation of \cite{yo} is also
substantially different from ours (there is a subtraction from the $\dot
g_{ij}$
equation, for instance, which we do not have, and the subtraction from the
$\dot K_{ij}$ equation is different from ours), though the coefficient of
subtraction from $\dot K_{ij}$ for this approach is small, comparable to
ours.
\bigskip

\subsection{Densitized Lapse}
\label{sec:densitized}

There is extensive evidence in the literature that a {\it densitized} lapse
improves the hyperbolicity\cite{gr-qc/0402123} of the (at best)
weakly hyperbolic ADM form of
the Einstein equations. We implement densitized lapse for single black hole
simulations by writing

\be
\alpha = \alpha_{\rm analytic} (g/g_{\rm analytic})^p,
\ee
where $\alpha_{\rm analytic}$ is the explicit lapse as a function of coordinates
given by \eref{eq:ksAlpha},
$g_{\rm analytic}$ is the analytic Kerr-Schild $3-$metric determinant
as a function of coordinates, ($g_{\rm analytic}=1+2H l_t^2$,
\cite{gr-qc/0002076}),
$g$ is the computational $3-$metric determinant,
and $p$ is an adjustable positive constant usually taken to be $\frac{1}{3}$
or
$\frac{1}{2}$. In fact we find that densitized lapse does not enhance
constraint-subtracted lifetime, though it does contribute substantially
to longer lifetime in {\it constrained} evolutions described below.

%{\bf insert  figures here}

\section{Constrained Evolution}

The evolution of Kerr-Schild data
must continue to satisfy the constraint equations,
Eqs.~(\ref{eq:constraintH})--(\ref{eq:constraintK})
as we evolve away from the initial data.  However,
if we choose nonoptimal (e.g. zero)
constraint subtraction for even nonspinning black holes, the evolution
leads to an eventual violation of the constraints. Hence we have
investigated constrained evolution, solving the constraint equations as part
of the time update of the evolution equations.

The postevaluated tracking of constraint errors [residual
postevaluation] shown in Figures
\ref{fig:standard_domain}--\ref{fig:vary_domain} used the direct
discretization of the constraint
equations \eref{eq:constraintH} and \eref{eq:constraintK}.
For constrained evolution we need instead to implement an accurate,
efficient method of constraint {\it solution}.
We adopt the conformal transverse-traceless method
of York and collaborators~\cite{YP}-\cite{Bowen+York} which consists of a
conformal decomposition with a scalar $\phi$ that adjusts the metric,
and a vector potential $w^i$ that adjusts the longitudinal components of the
extrinsic curvature.
The constraint equations are then solved for these new quantities  $\phi$,
$w^i$ such that
the complete solution fully satisfies the constraints.

Applying this approach to constrained evolution,
the metric and traceless extrinsic curvature in the middle of a timestep
(after an explicit integration forward in time)  are
taken as conformal trial functions  $\tilde{g}_{ij}$ and
$\tilde{A}^{ij}$.

The physical metric at the end of the full timestep
(i.e. after the constraint equation solve), $g_{ij}$,
and the trace-free part of the extrinsic
curvature at the end of the full timestep, $A_{ij}$,
are related to the background fields through a
conformal
factor:
\ba
g_{ij} &=& \phi^{4} \tilde{g}_{ij}, \label{confg1} \\
\label{confg}
A^{ij} &=& \phi^{-10} (\tilde{A}^{ij} + \tilde{(lw)}^{ij}).
\label{eq:conf_field}
\ea
Here $\phi$ is the conformal factor, and $\tilde{(lw)}^{ij}$
will be used to cancel any possible longitudinal contribution.
$w^i$ is a vector potential, and
\ba
\tilde{(lw)}^{ij} \equiv \tilde{\nabla}^{i} w^{j} + \tilde{\nabla}^{j} w^{i}
        - \frac{2}{3} \tilde{g}^{ij} \tilde{\nabla_{k}} w^{k}.
\label{lw}
\ea
The trace $K$ is not corrected:
\be
K = \tilde K.
\label{tk}
\ee
Writing the Hamiltonian and momentum constraint equations in terms of
the quantities in 
Eqs.~(\ref{confg1})--(\ref{tk}), we obtain four coupled
elliptic equations for the fields $\phi$ and $w^i$~\cite{YP}:
%\ba
%\tilde{\nabla}^2 \phi &=&  (1/8) \big( \tilde{R}\phi
%        + \frac{2}{3} \tilde{K}^{2}\phi^{5} -   \nonumber \\
%        & & \phi^{-7} (\tilde{A}{^{ij}} + (\tilde{lw})^{ij})
%            (\tilde{A}_{ij} + (\tilde{lw})_{ij}) \big),   \\
%\label{ell_eqPhi}
%\tilde{\nabla}_{j}(\tilde{lw})^{ij} &=& \frac{2}{3} \tilde{g}^{ij} \phi^{6}
%        \tilde{\nabla}_{j} K - \tilde{\nabla}_{j} \tilde{A}{^{ij}}.
%\label{ell_eqs}
%\ea
\begin{eqnarray}
\tilde{\nabla}^2 \phi &=&  \left(1/8\right) \left[ \tilde{R}\phi  \right.
       + \frac{2}{3} \tilde{K}^{2}\phi^{5} -   \nonumber \\
   & & \phi^{-7} \left(\tilde{A}{^{ij}} + (\tilde{lw})^{ij}\right)
     \left.  \left(\tilde{A}_{ij} + (\tilde{lw})_{ij} \right) \right], \label{ell_eqPhi} \\
     \tilde{\nabla}_{j}(\tilde{lw})^{ij} &=& \frac{2}{3} \tilde{g}^{ij} \phi^{6}
     \tilde{\nabla}_{j} \tilde{K} - \tilde{\nabla}_{j} \tilde{A}{^{ij}}. \label{ell_eqs}
\end{eqnarray}

These equations are solved to complete each time-update step.
The resulting solved $g_{ij}$ and $K_{ij}$ are taken as the data for the
next time-update. Notice that these equations require no specific gauge
choice. A similar approach also can be applied to other formulations which
generally have a larger number of constraints.

\subsection{Elliptic Equation Boundary Conditions}
\label{sec:boundary}

A solution of the elliptic constraint equations requires that boundary
data be specified on both the outer boundary {\em and } on the surfaces
of any masked regions.
For the elliptic solution here we can choose simple
conditions, $\phi = 1$ and $w^{i} = 0$, on the masked region surrounding
the singularity. Because we solve the problem
on a finite domain, we also must provide an outer boundary condition for
$\phi $ and $w^{i}$. For this demonstration of the technique, we
choose the same conditions at the outer boundary of the domain:
$\phi = 1$ and $w^{i} = 0$. In
long term evolution we expect the evolved solution to
converge (as the solution is refined) to a solution of the constraints,
so a global solution $\phi = 1$ and $w^{i} = 0$ is expected in this
analytic limit. For achievable resolutions, however,
the quantities $\phi $ and $w^{i}$ deviate from this prediction.

\section{Constrained Evolution Results}
\label{sec:physresults}

\begin{figure}
\begin{center}
\includegraphics[width=6.0in, angle=0]{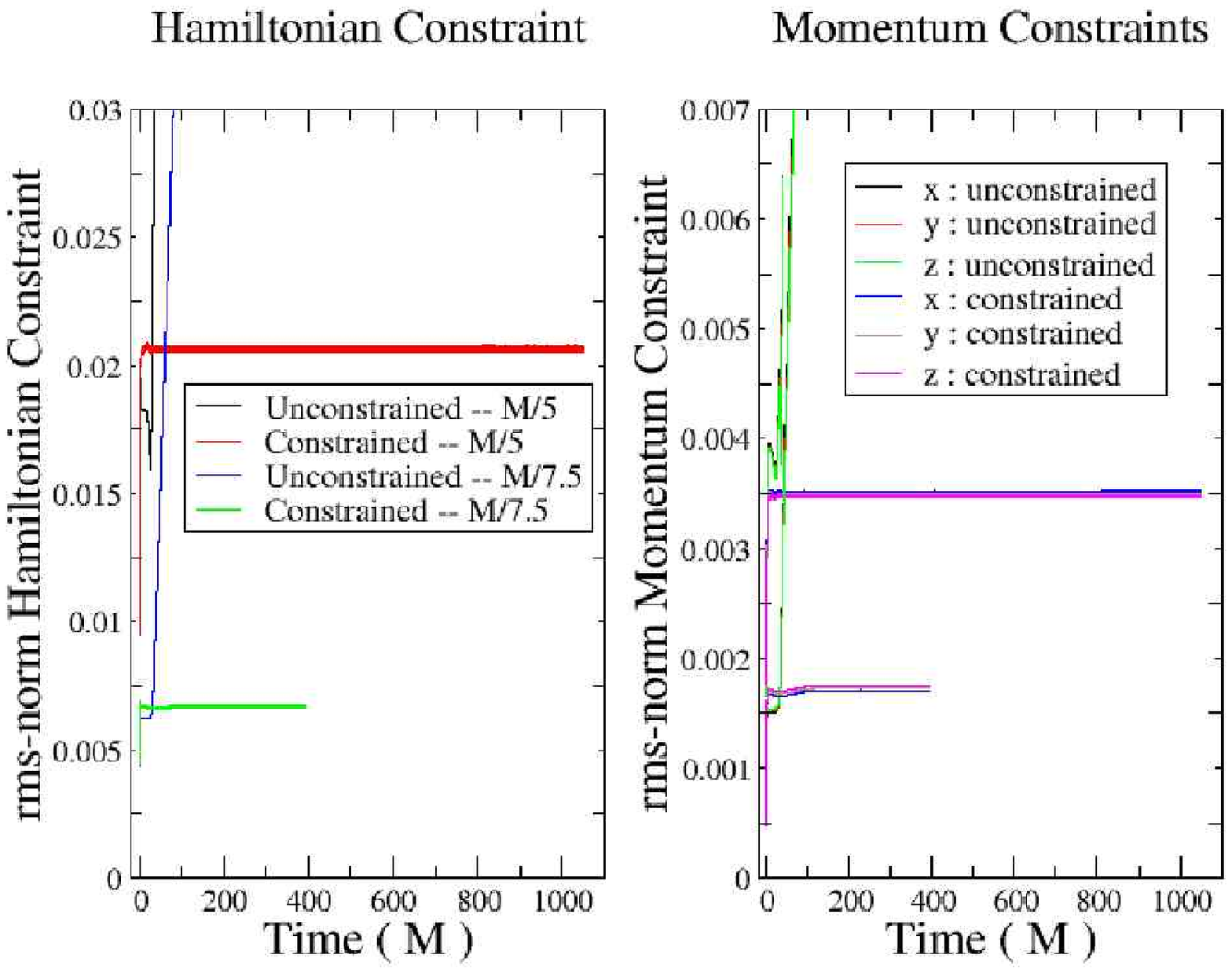}
\end{center}
\caption{The rms norm of the Hamiltonian and momentum constraints
for simulations of a Schwarzschild (nonspinning)
black hole with excision. None of the simulations used any constraint subtraction.
The simulations were performed at resolutions 
of either $M/5$ or $M/7.5$ on a domain size of $\pm 5M$. The long-lived runs employed
constrained evolution as described in the text. The short-lived runs were
unconstrained.
}
\label{fig:constrained}
\end{figure}

The elliptic constraint equations are solved either by a
PETSc \cite{PETSC_1,PETSC_2,PETSC_3} GMRES solver or
KINSOL \cite{SUNDIALS} GMRES solver;
the spatial differencing is fourth order. We present
below (Figures \ref{fig:constrained} - \ref{fig20.pdf})
the results of preliminary constrained evolution of
nonspinning black holes. Figure \ref{fig:constrained} shows the
rms norm of Hamiltonian and momentum constraints
for these simulations.
Compared 
to the relatively short term crash of the unconstrained
evolution, the constrained evolution clearly
does
stabilize single black hole evolutions in small domains, regardless of the
precise
subtraction. To fully understand the content of
Figure \ref{fig:constrained},
consider that, even
if the residual limit
in the solution of the constraint equations,
\eref{ell_eqPhi} -- \eref{ell_eqs}, is set
to extremely small values (it can be set very near to machine
precision, 
meaning that the discretized matrix form of equations
\eref{ell_eqPhi} --  \eref{ell_eqs}
can be solved to fractional errors of order $10^{-15}$), the
post-evaluations 
of the constraint residuals will typically show the
expected $zero$ only
to the internal discretization accuracy (here fourth order).
This is why 
the constrained solution shown in Figure \ref{fig:constrained}
shows a finite (but convergent) level of Hamiltonian
constraint violation.

Figures \ref{fig15.pdf}-\ref{fig20.pdf} show a constrained
$\dot g - \dot K$ evolution with densitized lapse, at three resolutions
($M/5, M/7.5, M/10$). The $M/5$ evolution became unstable before
$t= 100 M$. The $M/7.5$ run began showing large residuals at
$t \approx 350 M$. The $M/10$ run shows better behavior than $M/7.5$,
at least initially It shows a similar late time instability which
tracks (at smaller error) the behavior of the $M/7.5 $ case,
but around $150M$ ceases to be convergent; see Figure \ref{fig17.pdf}.
Figure  \ref{fig20.pdf} shows the 2d $z=0$ behavior of the residual
component 
$G_{xx}$ at $t = 100M$. The ``red-blue" pattern of the features
near the excision mask indicate that most of the error develops
there. The residual becomes more asymmetrical at
later times.
 
\section{CORRECTNESS OF CONSTRAINED EVOLUTION}
\label{sec:correctness}

The constraint maintenance approach uses what has been called in
magnetohydrodynamics, a {\it projection} method\cite{projection}.  This
method for constrained evolution raises questions about the meaning of
the solutions obtained. This is sometimes put bluntly: ``Accepting that
the method finds solutions of the full Einstein system, how do we know
that the found solution is the {\it right} one?" By this is meant that
the constraint solution step may somehow move the solution back to an
``erroneous" point on the space of constraint solutions. For instance,
it might be possible that although the evolution substep and the
constraint substep are individually convergent computational
processes, the result of combining them is in
some manner {\it not} convergent. (In the much
simpler MHD case there are analytical proofs that projection
minimizes the resultant error in the magnetic field, in a
convergent
way.\cite{projection})

\begin{figure}
\begin{center}
\includegraphics[width=5.0in, angle=0]{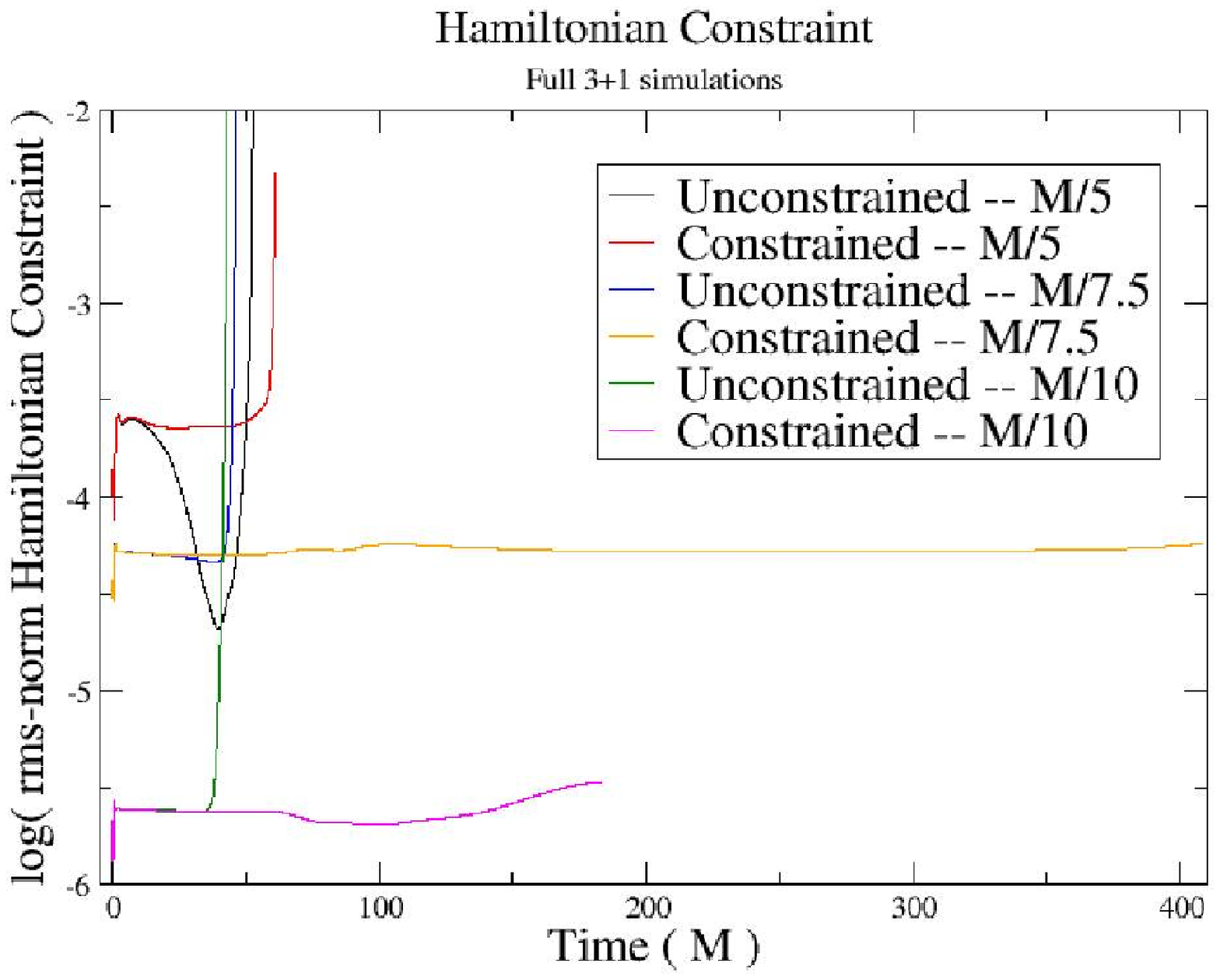}
\end{center}
\caption{The log of the rms norm for the Hamiltonian constraint
in simulations evolving a nonrotating Kerr-Schild black
hole with a spatial
domain of [$-10M . . . 10M$] at three resolutions:
$M/5, M/7.5,$ and $M/10.$ The excision
radius was $0.5M$ in all constrained and unconstrained
cases. All simulations
used a densitized lapse with $p = \frac{1}{3}$.
The constraints were
solved in the constrained
evolution cases every $0.05M$ everywhere on the domain
except those points where
$r < 2.0M$. Independent residual evaluations for the
constrained cases with
resolution M/7.5 and M/10 are found in Figure \ref{fig17.pdf}}
\label{fig15.pdf}
\end{figure}

\begin{figure}
\begin{center}
\includegraphics[width=5.0in, angle=0]{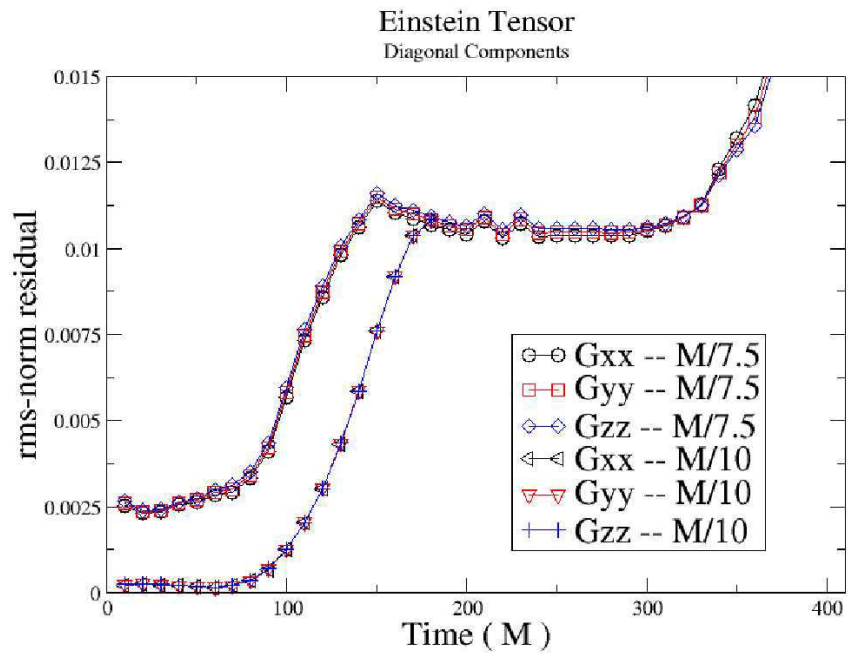}
\end{center}
\caption{The rms norm of the diagonal
spatial components of the Einstein tensor for the
M/7.5 and M/10 constrained simulations presented in
Figure \ref{fig15.pdf}}
\label{fig17.pdf}
\end{figure}

\begin{figure}
\begin{center}
\includegraphics[width=5.0in, angle=0]{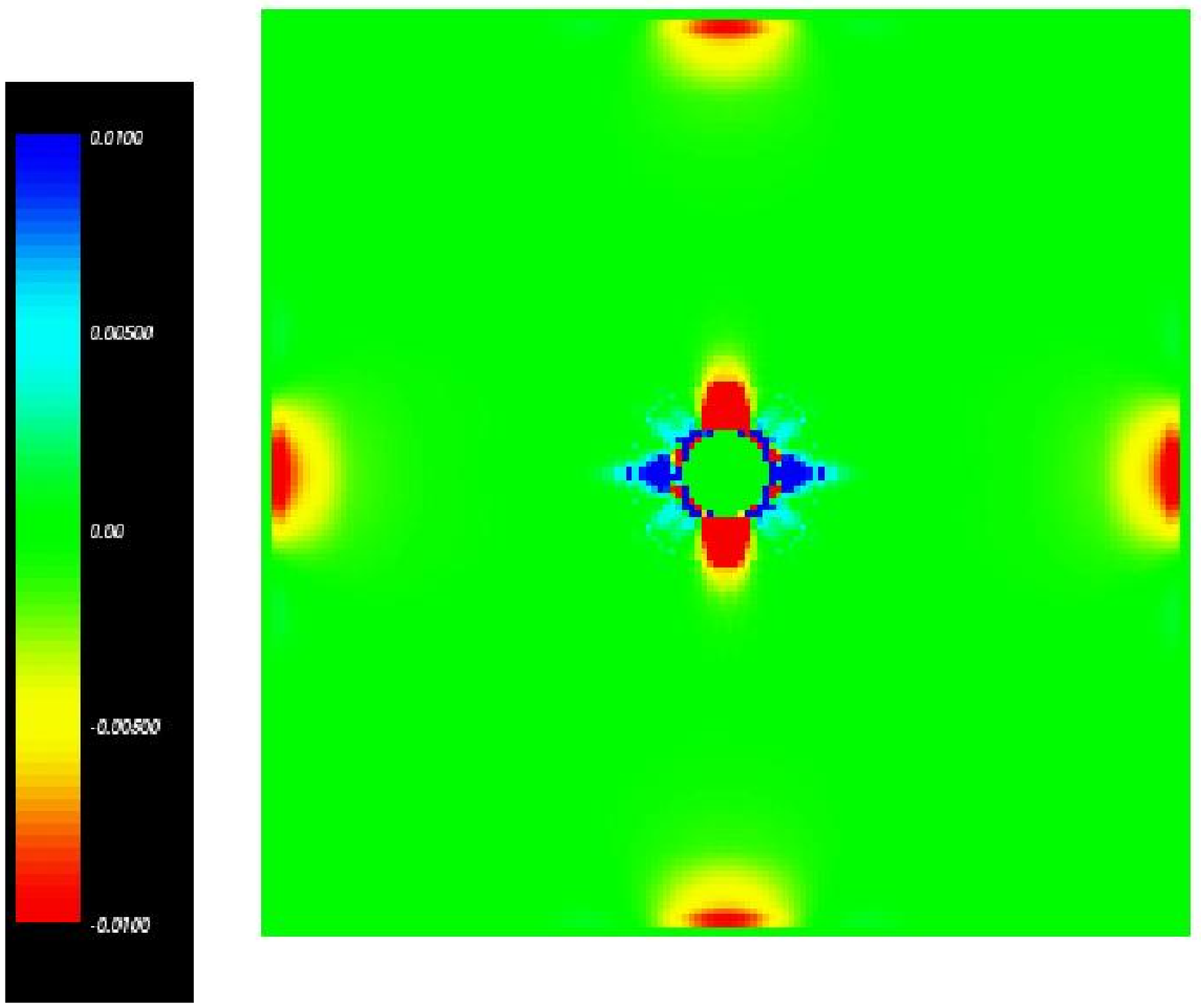}
\end{center}
\caption{The z = 0 plane of the
Einstein tensor component $G_{xx}$ at time
$100M$ for the $M/7.5$ constrained evolution
presented in Figures \ref{fig15.pdf}, \ref{fig17.pdf}.
}
\label{fig20.pdf}
\end{figure}

There are several parts to the response. (It will be clear that we do
not pretend 
to a rigorous analytical proof.)
To begin with, we have
constructed completely independent ``residual evaluators"
for the full Einstein system\cite{choptuikSB}. These evaluate the
Einstein tensor, working just from the metric produced by the
computational solution. They are completely different from the way
the equations are expressed in the constrained evolution code.
As we show in Figures \ref{fig15.pdf},\ref{fig17.pdf}, the
resulting residual is in every case initially small (order of
truncation error) and convergent. Thus we have
achieved a computational solution to the Einstein system.
Note that our
full Einstein equation residual evaluator
checks both the constraints (Einstein equations at one time),
and the evolution 
equations connecting different time steps.

The residual evaluators are written to return fourth order accurate
results. Since they have now been verified and show
convergence of the solution, we appeal
to the assumption that
the Einstein system is not {\it singular} at the solution
manifold. Thus we
expect that the computational result converges to the analytical
solution of the Einstein equations.  We have converged to a
spacetime configuration. Physical consistency and generality
imply that
it is the physically unique one that contains the initial
data slice. We also note that, as in the situation in Figure \ref{fig17.pdf},
the convergence 
is eventually lost in some simulations; these simulations are no longer
solving Einstein's equations.\\

It is of interest to ask why our approach has not been
implemented
previously.  A number of factors were at work. It has been
universally
assumed that the computational overhead of elliptic solvers is
excessive. Choptuik \cite{ChoptuikPC} indicates a time
penalty of $\times 2$, incurred by a fully constrained $2d$
evolution,
compared to free evolution. This cost is justified by the
much longer
physical lifetimes achieved in the ($2d$) evolutions of
Ref  \cite{gr-qc0301006}; Choptuik's factor of two in time
is considered a small
penalty. However, in previous implementations of
constrained ($2d$)
evolution, one additionally had the problem that the
solvers were 
restricted, for instance to conformally flat situations.
This required 
strong gauge constraints on the evolution, and meant that
generally one re-solved a strongly nonlinear equation
(comparable to the initial value problem), on each time step.
But we have found, even working with
straightforward package solvers, that the penalty for our
approach is only of order 30\%.
This is because our constraint solver (developed for Kerr-Schild
superposed initial data), is in fact completely general with no form
restrictions on the background. (There are, of course, general conditions
on 
the elliptic equations to allow their solution \cite{restrictions},
but we 
have encountered no difficulties in working with physically
realistic 
configurations). 

Thus our elliptic solver can use backgrounds
that are
strongly nonflat. They may be strongly nonflat, but they are
already very close to constraint solution.
This is so because they
arise from evolving one timestep with an accurate time
integrator; we use a package integrator which is typically
fourth order
accurate in time, and the spatial discretization
is also fourth order. There is thus only a very small
correction arising from the elliptic solve, and
minimal outer iteration is required. Further,
we take note of the fact that our explicit time
integration has a certain inherent order of truncation error.
Thus we do not in fact set the residual
limit for the elliptic solve anywhere near machine accuracy.
Instead we set it so that it produces errors which are consistent
in size with the
evolution truncation error.  This reduces the {\it internal}
iteration in the solver to a very small number.
Finally, computational resources are now becoming
adequate for constrained 3-d evolution.
$1$Tflop/sec Computers are now accessible,
making this work plausible. It
will always be the case that constrained evolution is more computation-
and memory- intensive than unconstrained, but the time has arrived that
interesting constrained evolutions are possible. Further, ongoing
computational infrastructure improvements will make this level of
computation generally accessible. Well within a decade, desktop access
to $10$ Tflop/sec will mean that constrained evolutions at LISA- or
LIGO- relevant resolutions will become unexceptional.

\section{Discussion}

We have demonstrated that the analytical formulation is not critical to long
term evolution of single black holes, if the correct constraint subtraction
is used.
Thus, ``modern" approaches that pose explicitly hyperbolic approaches are
not
essential; ``traditional" $\dot g$ - $\dot K$ methods produce comparably
long
evolutions. The evidence seems to be that there are many formulations and
subtraction schemes that lead to long-term single black hole stability; we
have found an especially simple one.

However, we have found, consistent with theoretical estimates, that the
precise subtraction is critical (a fraction to two or three decimals). To
attack this problem, we have carried out periodic solution of the elliptic constraint
equations as part of
the time integration, to enable fully constrained evolutions of
the Einstein equations. Our initial results demonstrate dramatic
improvement of long
term stability of a nonspinning black hole simulation. Because we solve the
constraints, constraint subtraction is irrelevant in this case. We are
beginning exploration of the constrained evolution
approach in spacetimes
involving single moving, and multiple interacting black holes. We find
substantial improvement from constraint solving in every simulation,
but we have not achieved infinite-lived $\dot g - \dot K$ simulations,
even with densitized lapse (which is known to improve the hyperbolicity
of the system of equations). We have begun an approach where
the inner excision is made at a constant coordinate surface that
coincides with the apparent horizon (essentially {\it spherical}, or
{\it spheroidal} coordinates) near the excision
region\cite{mattDiss}. This approach addresses concerns about the validity of the
stair-step (LEGO) excision region (\cite{gr-qc/0302072}).
(Additionally the best long-lived isolated black hole simulation to
date has been carried out by Ref \cite{schKidteuk} with a
pseudo-spectral method with apparent horizon conforming coordinates.)

Our approach uses coordinates which are spherical near the hole, with a
cartesian 
region further away. It is hoped that this approach may exhibit some of the
good 
properties and long lifetime of the codes described in \cite{schKidteuk}.

\section*{Acknowledgments}
Computations were performed at the Texas Advanced Computing Center
at the
University of Texas.  This work was supported by NSF grants PHY
0102204 and  PHY~0354842. Additionally, portions of this
work were conducted
at the Kavli Institute for Theoretical Physics,
The University of California at
Santa Barbara, under NSF grant PHY99 07947, and at the Laboratory for
High Energy Astrophysics, NASA/Goddard Space flight Center,
Greenbelt Maryland, with support from the University Space Research
Association.  
M.~Anderson acknowledges support from
a Department of Energy
Computational Science Graduate Fellowship administered by the Krell
Institute.

%This work is
%supported in part by NSF grants PHY~9800722, PHY~9800725, and
%PHY~0102204. Computations were performed at the NSF supercomputer
%center NCSA and the University of Texas AHPCC.


\newpage

\begin{thebibliography}{99}




\bibitem{schKidteuk} Mark A. Scheel, Lawrence E. Kidder, Lee Lindblom,
Harald P. Pfeiffer, Saul A. Teukolsky, ``Toward stable 3D numerical
evolutions of black-hole spacetimes," {\it Phys. Rev.} {\bf D66} 124005
(2002)[arXiv:gr-qc/0209115].


\bibitem{yo}Hwei-Jang Yo, Thomas W. Baumgarte, Stuart L. Shapiro, ``Improved
numerical stability of stationary black hole evolution calculations," {\it
Phys. Rev.} {\bf D66} 084026 (2002) [arXiv:gr-qc/0209066].





\bibitem{ADM} R.~Arnowitt, S.~Deser, and C.~Misner in Witten,
{\it Gravitation,
an Introduction to Current Research} (Wiley, New York 1962).

\bibitem{49Frittelli}Frittelli, S., and Reula, O., ``First-order
symmetric-hyperbolic Einstein equations with arbitrary fixed gauge'', {\it
Phys. Rev. Lett.}, {\bf 76}, 4667-4670, (1996).


%\bibitem{Matzner:1999pt}
%R.~Matzner, M.~F.~Huq and D.~Shoemaker,
%{\it Phys.\ Rev.}  {\bf D59}, 024015 (1999)[arXiv:gr-qc/9805023].



\bibitem{KerrSchild} R.~Kerr and A.~Schild, ``Some Algebraically Degenerate
Solutions of Einstein's Gravitational Field Equations,'' in
{\it Applications of Nonlinear Partial Differential Equations in
Mathematical Physics}, Proc. of Symposia B Applied Math., Vol XVII (1965);
``A New Class of Solutions of the Einstein Field Equations'',
{\it Atti del Congresso Sulla Relitivita Generale: Problemi Dell'Energia
E Onde Gravitazionala} G. Barbera, Ed. (1965).


\bibitem{SUNDIALS} A.~Hindmarsh, R.~Serban and C.~Woodward, ``SUNDIALS home
page",
\textit{http://www.llnl.gov/CASC/sundials/} (2002).

\bibitem{CVODES} A.~Hindmarsh and R. Serban, ``User Documentation for
CVODES",
(Center for Applied Scientific Computer, 
Lawrence Livermore National Laboratory, 2002).

\bibitem{KINSOL} A.~Taylor and A. Hindmarsh, ``User Documentation for
{KINSOL}, A Nonlinear Solver for Sequentail and Parallel Computers",
(Center for Applied Scientific Computer, 
Lawrence Livermore National Laboratory, 1998).

\bibitem{gr-qc/0402123}Gabriel Nagy, Omar E. Ortiz and Oscar A. Reula
``Strongly hyperbolic second order Einstein's evolution equations",
gr-qc/0402123 (2004).

\bibitem{YP} J.~York and  T.~\ Piran
        ``The Initial Value Problem and Beyond'',
        \textit{Spacetime and Geometry: The Alfred Schild
        Lectures}, R.~Matzner and L.~Shepley Eds.
        University of Texas Press, Austin, Texas. (1982);
        G.~Cook, ``Initial Data for the Two-Body Problem
        of General Relativity'', Ph.D. Dissertation, The University
        of North Carolina at Chapel Hill (1990).


\bibitem{MY} N.~{\'O}.~Murchadha and J.~W.~York, Jr.,
                   {\it Phys.\ Rev.}  {\bf D10}, 428 (1974);
             N.~{\'O}.~Murchadha and J.~W.~York, Jr.,
                   {\it Phys.\ Rev.} {\bf D10}, 437 (1974);
             N.~{\'O}.~Murchadha and J.~W.~York, Jr.,
                   {\it Gen.\ Relativ.\ Gravit.} {\bf 7} 257 (1976).

\bibitem{York} J.~W.~York, Jr., {\it Phys.\ Rev.\ Lett.} {\bf 82},
        1350 (1999).
        


\bibitem{Mathews} J.~R.~Wilson and G.~J.~Mathews,
    {\it Phys.~Rev.~Lett.} {\bf 75}, 4161 (1995);
    J.~R.~Wilson and G.~J.~Mathews, and P.~Marronetti,
    {\it Phys.~Rev.} {\bf D54}, 1317 (1996)





\bibitem{Bowen+York} J.~Bowen and J.~W.~York,
                     {\it Phys. Rev.} {\bf D21}, 2047 (1980).

\bibitem{gr-qc/0002076}
Mijan F. Huq, Matthew W. Choptuik and Richard A. Matzner,
``Locating Boosted Kerr and Schwarzschild Apparent Horizons"
{\it Physical Review} {\bf D66},~084024~(2002).

\bibitem{ChoptuikPC} M.~Choptuik, Personal communication.

\bibitem{PETSC_1} S.~Balay, K.~Buschelman, W.~D.~Gropp, D.~Kaushik,
M.~Knepley, L.~C.~McInnes, B.~F.~Smith and H.~Zhang, ``PETSc home page",
\textit{http://www.mcs.anl.gov/petsc} (2001).

\bibitem{PETSC_2} S.~Balay, K.~Buschelman, W.~D.~Gropp, D.~Kaushik,
M.~Knepley, L.~C.~McInnes, B.~F.~Smith and H.~Zhang, ``PETSc Users Manual",
\textit{ANL-95/11 - Revision 2.1.5} (Argonne National Laboratoy, 2002).

\bibitem{PETSC_3} S.~Balay, K.~Buschelman, W.~D.~Gropp, D.~Kaushik,
M.~Knepley, L.~C.~McInnes, B.~F.~Smith and H.~Zhang, ``Efficient Management
of 
Parallelism in Object Oriented Numerical Software Libraries",
\textit{Modern Software Tools in Scientific Computing}, E.~Arge,
A.~M.~Bruaset, and H.~P.~Langtangen, editors,
(Birkhauser Press, 1997), pp. 163-202.

\bibitem{gr-qc/0302072} Gioel Calabrese, Luis Lehner, David Neilsen,
Jorge Pullin, Oscar Reula, Olivier Sarbach, Manuel Tiglio,
``Novel finite-differencing techniques for numerical relativity:
application to black hole excision", {\it Classical and Quantum Gravity}
{\bf 20} L245-L252 (2003).

\bibitem{projection}G. Toth ``The ${\bf \nabla \cdot}$ {{\bf B}} $= 0$ Constraint in Shock-Capturing
Magnetohydrodynamics Codes"
{\it Jour. Comp. Phys.} {\bf 161} 605 (2002).

\bibitem{choptuikSB} Personal communication (2003).


\bibitem{gr-qc0301006}
M. W. Choptuik,  E. W. Hirschmann,  S. L. Liebling,  and F. Pretorius,
``An Axisymmetric Gravitational Collapse Code"
 {\it Class. Quant. Grav.} {\bf 20}, 1857-1878 (2003).

\bibitem{restrictions} Y. Choquet-Bruhat, J. Isenberg and J. W. York,
{\it Phys. Rev.} {\bf D61} 084034 2000

\bibitem{mattDiss}Matthew Anderson, ``Constrained Evolution in Numerical Relativity"
Ph.D. dissertation, The University of Texas at Austin (2004).

\end{thebibliography}
\end{document}